\documentclass[11pt,fleqn]{article}

\usepackage{a4}
\usepackage{amstext}
\usepackage{amsfonts}
\usepackage{amssymb}
\usepackage{color}
\usepackage{epsfig}
\usepackage{axodraw}

\newcommand{\ltapprox}{\raisebox{-0.5ex}{$\,\stackrel{<}{\scriptstyle\sim}\,$}}

\newcommand{\bea}{\begin{eqnarray}}
\newcommand{\eea}{\end{eqnarray}}
\newcommand{\beq}{\begin{equation}}
\newcommand{\beqns}{\begin{eqnarray*}}
\newcommand{\eeqns}{\end{eqnarray*}}
\newcommand{\eeq}{\end{equation}}

\newcommand{\pdir}{p\kern -5.2pt\raise 0.2ex\hbox {/}}
\newcommand{\vdir}{v\kern -5.75pt\raise 0.15ex\hbox {/}}
\newcommand{\kdir}{k\kern -5.75pt\raise 0.15ex\hbox {/}}
\newcommand{\epsdir}{\epsilon\kern -5.0pt\raise 0.15ex\hbox {/}}
\newcommand{\bvdir}{\bar{v}\kern -5.75pt\raise 0.15ex\hbox {/}}
\newcommand{\Ddir}{D\kern -7.75pt\raise 0.20ex\hbox {/}}
\newcommand{\Adir}{A\kern -7.75pt\raise 0.20ex\hbox {/}}
\newcommand{\ldir}{l\kern -5.0pt\raise 0.2ex\hbox{/}}
\newcommand{\varepsdir}{\varepsilon\kern -5.5pt\raise 0.15ex\hbox{/}}

\newcommand{\dlz}{\stackrel{\leftarrow}{D^0}}
\newcommand{\drz}{\stackrel{\rightarrow}{D^0}}

\newcommand{\lgl}{\langle}
\newcommand{\rgl}{\rangle}

\def\elematrice#1#2#3{\lgl#1|#2|#3\rgl}

\begin{document}


\begin{flushright}
DESY 09-033 \\ SFB/CPP-09-26 \\ LPT-ORSAY 09-17 \\ HU-EP-09/12
\end{flushright}

\begin{center}

{\huge \bf Lattice calculation of the Isgur-Wise}

{\huge \bf functions $\tau_{1/2}$ and $\tau_{3/2}$ with dynamical quarks}

\vspace{0.5cm}

\textbf{Benoit Blossier}$^{a,b}$, \textbf{Marc Wagner}$^{c}$, \textbf{Olivier P\`ene}$^{b}$  \\

{\par\centering \vskip 0.7 cm\par}
{\par\centering \textsl{$^a$ 
DESY, Platanenallee 6, D-15738 Zeuthen, Germany}}
{\par\centering \vskip 0.3 cm\par}
{\par\centering \textsl{$^b$
Laboratoire de Physique Th\'eorique (B\^at.210), Universit\'e
Paris-Sud XI,\\
Centre d'Orsay, 91405 Orsay-Cedex, France.}} \\
{\par\centering \vskip 0.3 cm\par}
{\par\centering \textsl{$^c$ 
Humboldt-Universit\"at zu Berlin, Institut f\"ur Physik, Newtonstra{\ss}e 15, D-12489 Berlin, Germany  \\
}
}

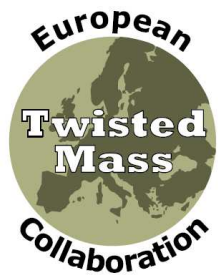
\vspace{0.5cm}

\begin{picture}(0,0)%
\includegraphics{Logo.pstex}%
\end{picture}%
\setlength{\unitlength}{4144sp}%
\begingroup\makeatletter\ifx\SetFigFont\undefined%
\gdef\SetFigFont#1#2#3#4#5{%
  \reset@font\fontsize{#1}{#2pt}%
  \fontfamily{#3}\fontseries{#4}\fontshape{#5}%
  \selectfont}%
\fi\endgroup%
\begin{picture}(1620,1620)(1,-781)
\end{picture}%

\vspace{0.4cm}

March 13, 2009

\end{center}

\vspace{0.1cm}

\begin{tabular*}{16cm}{l@{\extracolsep{\fill}}r} \hline \end{tabular*}

\vspace{-0.4cm}
\begin{center} \textbf{Abstract} \end{center}
\vspace{-0.4cm}

We perform a dynamical lattice computation of the Isgur-Wise functions $\tau_{1/2}$ and $\tau_{3/2}$ at zero recoil. We consider three different light quark masses corresponding to $300 \, \textrm{MeV} \ltapprox m_\mathrm{PS} \ltapprox 450 \, \textrm{MeV}$, which allow us to extrapolate our results to the physical $u/d$ quark mass. We find \\ $\tau_{1/2}(1) = 0.296(26)$ and $\tau_{3/2}(1) = 0.526(23)$. Uraltsev's sum rule is saturated up to $80 \%$ by the ground state.  We discuss implications regarding semileptonic decays $B \rightarrow X_c \, l \, \nu$ and the associated ``$1/2$ versus $3/2$'' puzzle.

\begin{tabular*}{16cm}{l@{\extracolsep{\fill}}r} \hline \end{tabular*}

\thispagestyle{empty}


\newpage

\setcounter{page}{1}

\section{Introduction}

The semileptonic decay of $B$ mesons into positive parity charmed mesons (often
referred to as $D^{**}$'s) is an important and debated issue. Important,
because no accurate measurement of the $V_{c b}$ CKM angle will be possible, if
these channels, which represent about one quarter of the semileptonic decays,
are not well understood. Debated, because there seems to be a persistent
discrepancy between claims from theory and from experiment \cite{Bigi:2007qp}.

Two types of $D^{**}$'s are seen, two ``narrow resonances'' and a couple of ``broad resonances'', grossly speaking in the same mass region. While experiments point towards a dominance of the broad resonances in semileptonic decays, theory, when using the heavy quark limit, points rather towards a dominance of the narrow resonances. To clarify the situation ref.~\cite{Bigi:2007qp}
called for actions on both the experimental and the theoretical side.

The theoretical argument relies on a series of sum
rules \cite{LeYaouanc:1996bd,Uraltsev:2000ce} derived from QCD comforted by model calculations \cite{Morenas:1997nk,Ebert:1998km,Ebert:1999ga}. Lattice calculations are needed
to give a more quantitative prediction stemming directly from QCD. A preliminary computation was performed in \cite{Becirevic:2004ta}, but only in quenched QCD and with a
marginal signal-to-noise ratio. In this letter we report on the first unquenched
computation using $N_f = 2$ flavor gauge configurations with Wilson twisted quarks
generated by the European Twisted Mass Collaboration (ETMC). The spectrum of
heavy-light mesons in the static limit has already been reported \cite{Jansen:2008ht,Jansen:2008si}.


\subsection{Spectrum in the heavy quark limit}

We treat both $b$ and $c$ quarks via static Wilson lines, i.e.\
consider their infinite mass limit. In this limit the meson spectrum is
constructed by combining the spin $1/2$ of the heavy quark with the total
angular momentum and parity $j^\mathcal{P}$ of the light degrees of freedom (light quarks and gluons) \cite{Isgur:1989vq,Isgur:1989ed,Isgur:1990jf}. The two
lightest negative parity mesons $B$ and $B^*$ (or $D$ and $D^*$) are degenerate and described by the same $S \equiv (1/2)^-$ state of light particles. The lightest (non-radially excited) positive parity states can be decomposed into two
degenerate doublets: $P_- \equiv (1/2)^+$ and $P_+ \equiv (3/2)^+$. The
total angular momenta $J^\mathcal{P}$ of the $P_-$ ($P_+$)
mesons are $0^+$, $1^+$ ($1^+$, $2^+$). The mixing between the two $1^+$
states is suppressed in the heavy quark limit.

It is generally believed that
the narrow (broad) resonances are of the $P_+$ ($P_-$) type, since in the
heavy quark limit they decay into $D^{(*)} \pi$ via a $D$ ($S$) wave.
The $D$ wave decays are supposed to be suppressed by a centrifugal
barrier, if the final state momenta are not too large. 


\subsection{Decay form factors in the heavy quark limit}

In the heavy quark limit the semileptonic decay of a pseudoscalar meson into
$D^{**}$ is governed by only  two form factors \cite{Isgur:1990jf},
$\tau_{1/2}(w)$ and $\tau_{3/2}(w)$, where  $w \equiv v_B \cdot v_{D^{**}} \geq
1$ with $v_B$ and $v_{D^{**}}$ denoting the four-velocity of heavy-light meson $H$ being defined by  $v_H \equiv p_H / m_H$. Uraltsev has proven the following sum rule \cite{Uraltsev:2000ce}:
\begin{eqnarray}
\label{EQN_uraltsev} \sum_n \Big|\tau_{3/2}^{(n)}(1)\Big|^2 -
\Big|\tau_{1/2}^{(n)}(1)\Big|^2 \ \ = \ \ \frac{1}{4} ,
\end{eqnarray}
where $\tau_{3/2}^{(n)}(w)$ ($\tau_{1/2}^{(n)}(w)$), $n = 0,\ldots,\infty$
are the form factors for the decay into  the $P_+$ ($P_-$) meson and the
tower of its radial excitations\footnote{By definition  $\tau_j(w) \equiv
\tau_j^{(0)}(w)$, $j = 1/2, 3/2$.}. $w = 1$ corresponds to the zero recoil
situation,  i.e.\ the $B$ and the $D^{**}$ meson have the same
velocity. Eqn.\ (\ref{EQN_uraltsev}) is one of the major among many
theoretical arguments in favor of the narrow resonance
dominance~\cite{Bigi:2007qp}. 
 
Our goal in this paper is to make a direct lattice calculation of
$\tau_{1/2}(1)$ and $\tau_{3/2}(1)$  using static quarks represented by
Wilson lines \cite{Eichten:1989zv}. However, there is the  problem that the
$B \rightarrow D^{**}$ decay amplitude is suppressed at $w = 1$ due
to vanishing kinematical  factors, which multiply $\tau_j(1)$. This is also
a centrifugal barrier effect, i.e.\ it is impossible to give  angular
momentum to a meson at rest. Consequently, a computation of the weak current matrix
element will trivially give zero.  To overcome this difficulty, we use a
method, which amounts to compute the operator matrix element based on an 
expression of the derivative of that matrix element  in terms of the recoil
four-velocity of the final meson \cite{Leibovich:1997em,Becirevic:2004ta}.
Thanks to the translational invariance in time of the heavy quark Lagrangian this 
is then proportional to $\tau_j(1) (m_{H^j} - m_H)$, $j = 1/2, 3/2$  (cf.\ eqns.\
(\ref{EQN631}) and (\ref{EQN632})). The mass splittings $m_{H^{*
*}} - m_H$ have already  been computed in the static limit with
precisely the same setup we are using in this paper 
\cite{Jansen:2008ht,Jansen:2008si}, i.e.\ by using $N_f = 2$ ETMC gauge
configurations. We are thus in a  position to compute $\tau_{1/2}(1)$ and
$\tau_{3/2}(1)$ and to confront it with the Uraltsev and other sum  rules
as well as with other non-perturbative estimates (QCD sum
rules, quark models).

Our work should help to clarify the situation in the heavy quark limit. A
fair comparison with experiment  further needs to estimate the systematic
error stemming from the heavy quark limit. After all, the charm  quark is
not so heavy. The authors of \cite{Ebert:1998km,Ebert:1999ga} argue that
large $\mathcal{O}(1/m_Q)$ corrections are present. This issue can also be
addressed by lattice QCD, but in this work we restrict our computations to the
static limit.

The paper is organized as follows. In section~\ref{SEC_basics} we recall the method used to
compute $\tau_{1/2}(1)$ and $\tau_{3/2}(1)$.  In
section~\ref{SEC_computation} we report on the lattice calculation of
$\tau_{1/2}(1)$ and $\tau_{3/2}(1)$. In section \ref{SEC_renorm} we
perturbatively compute the renormalization constant of the heavy-heavy
current and we conclude in section \ref{SEC_concl}.



\section{\label{SEC_basics}Principle of the calculation}

To compute the zero-recoil Isgur-Wise functions $\tau_{1/2}(1)$ and $\tau_{3/2}(1)$ by means of lattice QCD, we use a method proposed in 
\cite{Becirevic:2004ta}. We remind it here just for comfort of the reader.

The method consists in using a series of relations derived in 
ref.~\cite{Leibovich:1997em}. With $v' = (1,0,0,0)$
and $v = v' + v_{\perp}$ denoting the velocities of the ingoing and outgoing mesons, where 
$v_\perp$ is spatial up
to higher orders in the difference $v'-v$, we 
assume that for some Dirac matrix $\Gamma_l$
\begin{eqnarray}
\label{tau1} \langle H^{**}(v') | \bar Q(v') \Gamma_l Q(v) |  H^{(*)}(v) \rangle \ \ = \ \ t_l^m v_{\perp m} \tau_j(w) + \cdots
\end{eqnarray}
Here $w \equiv v \cdot v'$, $j=1/2$, $3/2$ and $l,m=1,2,3$ are spatial
indices. $t_l^m$ is a tensor, which depends on the final state
($H^{**}$) and the initial state ($H^{*}$ or $H$), and $Q(v)$ is the
static quark field in Heavy Quark Effective Theory. The dots
represent higher order terms in  $v'-v$. From translational invariance in
time direction,
\begin{eqnarray}
\nonumber & & \hspace{-0.7cm} -i \partial_0 \langle H^{**}(v') | \bar Q(v') \Gamma_l Q(v) | H^{(*)}(v) \rangle \\
\nonumber & & = \ \ -i \langle H^{**}(v') | \bar Q(v') \Big[\Gamma_l \drz +
\dlz \Gamma_l\Big] Q(v) |  H^{(*)}(v) \rangle \\
\label{D0} & & = \ \ t_l^m v_{\perp m} \tau_j(w) \Big(m_{H^{**}} - m_{H}\Big) + \cdots
\end{eqnarray}
Then we use the field equation $(v \cdot D) Q(v) = 0$:
\begin{eqnarray}
D^0 Q(v') \ \ = \ \ 0 \quad ,\quad D^0 Q(v) \ \ = \ \ -(D \cdot v_\perp) Q(v) ,
\end{eqnarray}
 whence from eqn.\ (\ref{D0}) 
\begin{eqnarray}
\label{voila} i \langle H^{**}(v') | \bar Q(v') \Gamma_l (D \cdot v_\perp) Q(v) | H^{(*)}(v) \rangle \ \ = \ \ t_l^m v_{\perp m} \tau_j(w) \Big(m_{H^{**}} - m_{H}\Big) + \cdots ,
\end{eqnarray}
which, in the limit $v_\perp \to 0$, converges to the relation
\begin{eqnarray}
i \langle H^{**}(v) | \bar Q(v) \Gamma_l D^m Q(v) | H^{(*)}(v) \rangle \ \ = \ \ t_l^m \tau_j(1) \Big(m_{H^{**}} - m_{H}\Big) .
\end{eqnarray}
Applying eqn.\ (\ref{tau1}) to the $J = 0$ $H^*_0$ state we get
from ref.~\cite{Isgur:jf}
\begin{eqnarray}
\label{tau0} \langle H^*_0 (v') | A_i | H(v) \rangle \ \ \equiv \ \ -\tau_{1/2}(w) v_{\perp i} ,
\end{eqnarray}
where $A_i$ is the axial current in spatial direction $i$, and where the normalization of the states is $1 / \sqrt{2 m}$ times the one used in ref.~\cite{Isgur:jf}. From eqn.~(\ref{tau0}) follows
\begin{eqnarray}
\label{scalaire} \langle H^*_0(v) | A_i D_j | H(v) \rangle \ \ = \ \ i g_{i j}\left (m_{H^*_0} - m_{H}\right) \tau_{1/2}(1). 
\end{eqnarray}
Analogously for the $J = 2$ $H^*_2$ state we have
\begin{eqnarray}
\langle H^*_2 (v') | A_i | H(v) \rangle \ \ \equiv \ \ \sqrt{3} \tau_{3/2}(w) \epsilon^{* j}_i v_{\perp j} + \cdots ,
\end{eqnarray}
where $\epsilon^{* j}_i$ is the polarization tensor, whence
\begin{eqnarray}
\label{tenseur} \langle H^*_2(v) | A_i D_j | H(v) \rangle \ \ = \ \ -i \sqrt{3}\Big(m_{H^*_2} - M_H\Big) \tau_{3/2}(1) \epsilon^*_{i j} .
\end{eqnarray}
Finally $\tau_{1/2}(1)$ and $\tau_{3/2}(1)$  can be obtained from the following
matrix elements:
\begin{eqnarray}
\label{EQN631} & & \hspace{-0.7cm} \tau_{1/2}(1) \ \ = \ \ \bigg|\frac{\langle
H_0^* | \bar{Q} \gamma_5 \gamma_z D_z Q | H \rangle}{m_{H_0^*} - m_{H}}
\bigg| \\
\label{EQN632} & & \hspace{-0.7cm} \tau_{3/2}(1) \ \ = \ \ \bigg|\frac{\langle
H_2^* | \bar{Q} \gamma_5 (\gamma_x D_x - \gamma_y D_y) Q | H
\rangle}{\sqrt{6} (m_{H_2^*} - m_{H})} \bigg| .
\end{eqnarray}

There is no mixing of the operators $A_i D_j$ with dimension 3 (hence linearly divergent)
heavy-heavy operators to be feared on the lattice: indeed we are interested in a 
parity-changing transition and all dimension 3 operators have vanishing matrix elements
between positive and negative parity states\footnote{Of course the situation is different by
instance for the matrix element $\elematrice{H}{\bar{h}{\bf D}^2 h}{H}$, related to the 
HQET parameter $\lambda_1$ or the kinetic momentum $\mu^2_\pi$ for which a subtraction is
necessary to its computation on the lattice \cite{CrisafulliPG, DellaMorteCB}.}. There are
no logarithmic divergence either thanks to the vanishing of the vector and axial currents'
anomalous dimension in HQET at zero recoil. By consequence there is no conceptual issue
concerning the extrapolation to the continuum limit of such a calculation. It needs only a
finite renormalization constant to match the lattice result with a continuum-like scheme
value, as we will discuss in Section \ref{SEC_renorm}.



\section{\label{SEC_computation}Lattice computation of $\tau_{1/2}$ and
$\tau_{3/2}$ at zero recoil}


\subsection{\label{SEC_setup}Simulation setup}

We use $N_f = 2$ flavor $24^3 \times 48$ Wilson twisted mass gauge configurations produced by the European Twisted Mass Collaboration (ETMC). Here we only give a brief summary of the setup, which is explained in detail in \cite{Boucaud:2007uk,Urbach:2007rt,Boucaud:2008xu}.

The gauge action is tree-level Symanzik improved \cite{Weisz:1982zw} with $\beta = 3.9$ corresponding to a lattice spacing $a = 0.0855(5) \, \textrm{fm}$:
\begin{eqnarray}
S_\mathrm{G}[U] \ \ = \ \ \frac{\beta}{6} \bigg(b_0 \sum_{x,\mu\neq\nu} \textrm{Tr}\Big(1 - P^{1 \times 1}(x;\mu,\nu)\Big) + b_1 \sum_{x,\mu\neq\nu} \textrm{Tr}\Big(1 - P^{1 \times 2}(x;\mu,\nu)\Big)\bigg) ,
\end{eqnarray}
where $b_0 = 1 - 8 b_1$ and $b_1 = -1/12$.

The fermionic action is Wilson twisted mass with two degenerate flavors \cite{Frezzotti:2000nk,Frezzotti:2003ni,Shindler:2007vp}:
\begin{eqnarray}
S_\mathrm{F}[\chi,\bar{\chi},U] \ \ = \ a^4 \sum_x \bar{\chi}(x) \Big(D_{\rm W} + i\mu_\mathrm{q}\gamma_5\tau_3\Big) \chi(x) , 
\end{eqnarray}
where
\begin{eqnarray}
\label{eq:Wilson} D_\mathrm{W} \ \ = \ \ \frac{1}{2} \Big(\gamma_\mu \Big(\nabla_\mu + \nabla^*_\mu\Big) - a \nabla^*_\mu \nabla_\mu\Big) + m_0 ,
\end{eqnarray}
$\nabla_\mu$ and $\nabla^*_\mu$ are the standard gauge covariant forward and backward derivatives, $m_0$ and $\mu_\mathrm{q}$ are the bare untwisted and twisted quark masses and $\chi = (\chi^{(u)} \, , \, \chi^{(d)})$ are the fermionic fields in the twisted basis.

We consider three different values of the light quark mass, which amount to ``pion masses'' in the range $300 \, \textrm{MeV} \ltapprox m_\mathrm{PS} \ltapprox 450 \, \textrm{MeV}$ (cf.\ Table~\ref{TAB001}). $m_0$ has been tuned to its critical value at the lightest $\mu_\mathrm{q}$ value, i.e.\ at $\mu_\mathrm{q} = 0.0040$.



\begin{table}[h!]
\begin{center}
\begin{tabular}{|c|c|c|}
\hline
 & & \vspace{-0.40cm} \\
$\mu_\mathrm{q}$ & $m_\mathrm{PS}$ in MeV & number of gauge configurations \\
 & & \vspace{-0.40cm} \\
\hline
 & & \vspace{-0.40cm} \\
$0.0040$ & $314(2)$ & $1400$ \\
$0.0064$ & $391(1)$ & $1450$ \\
$0.0085$ & $448(1)$ & $1350$\vspace{-0.40cm} \\
 & & \\
\hline
\end{tabular}
\caption{\label{TAB001}twisted quark masses $\mu_\mathrm{q}$, pion masses $m_\mathrm{PS}$ and number of gauge configurations.}
\end{center}
\end{table}


\subsection{Static and light quark propagators}

The propagator of a static quark is essentially a Wilson line in time direction:
\begin{eqnarray}
\Big\langle Q(x) \bar{Q}(y) \Big\rangle_{Q,\bar{Q}} \ \ = \ \ \delta^{(3)}(\mathbf{x}-\mathbf{y}) U^{(\textrm{HYP2})}(x;y) \bigg(\Theta(y_0-x_0) \frac{1 - \gamma_0}{2} + \Theta(x_0-y_0) \frac{1 + \gamma_0}{2}\bigg) ,
\end{eqnarray}
where $\langle \ldots \rangle_{Q,\bar{Q}}$ denotes the integration over the static quark field and $U(x;y)$ is a path ordered product of links along the straight path from $x$ to $y$. To improve the signal-to-noise ratio we use the HYP2 static action \cite{Hasenfratz:2001hp,DellaMorte:2003mn,Della Morte:2005yc}.

For the light quarks we use four stochastic spin diluted timeslice propagators ($\mathcal{Z}_2 \times \mathcal{Z}_2$ sources with randomly chosen components $\pm 1 \pm i$) for each gauge configuration. For details we refer to \cite{Jansen:2008si}, where exactly the same setup has been used.


\subsection{Static-light meson creation operators}

In the static limit there are no interactions involving the heavy quark spin. Therefore, it is convenient to classify static-light mesons according to $j^\mathcal{P}$, where $j$ denotes the angular momentum of the light degrees of freedom and $\mathcal{P}$ parity. In particular we are interested in the sectors $j^\mathcal{P} = (1/2)^-$, $j^\mathcal{P} = (1/2)^+$ and $j^\mathcal{P} = (3/2)^+$. We label the corresponding static-light mesons, i.e.\ the ground states in these angular momentum/parity sectors, by $S$, $P_-$ and $P_+$ respectively.

To create such static-light mesons on the lattice we use operators
\begin{eqnarray}
\mathcal{O}^{(\Gamma)}(\mathbf{x}) \ \ = \ \ \bar{Q}(\mathbf{x}) \sum_{\mathbf{n} = \pm \hat{\mathbf{e}}_1 , \pm \hat{\mathbf{e}}_2 , \pm \hat{\mathbf{e}}_3} \Gamma(\hat{\mathbf{n}}) U(\mathbf{x};\mathbf{x}+r \mathbf{n}) \chi^{(u)}(\mathbf{x}+r \mathbf{n}) ,
\end{eqnarray}
where $\bar{Q}$ creates a static antiquark at position $\mathbf{x}$, $\chi^{(u)}$ creates a light quark in the twisted basis at position $\mathbf{x}+r \mathbf{n}$, $U$ is a product of spatial links along the straight path between $\mathbf{x}$ and $\mathbf{x}+r \mathbf{n}$, and $\Gamma$ is a combination of spherical harmonics and $\gamma$ matrices yielding a well defined behavior under cubic rotations (cf.\ Table~\ref{TAB002}).

\begin{table}[h!]
\begin{center}

\begin{tabular}{|c||c|c|}
\hline
 & & \vspace{-0.40cm} \\
$\Gamma(\hat{\mathbf{n}})$ & $\mathrm{O}_\mathrm{h}$ & $j$ \\
 & & \vspace{-0.40cm} \\
\hline
 & & \vspace{-0.41cm} \\
\hline
 & & \vspace{-0.40cm} \\
$\gamma_5$ & $A_1$ & $1/2 \ , \ 7/2 \ , \ ...$ \\
 & & \vspace{-0.40cm} \\
$1$ & & $1/2 \ , \ 7/2 \ , \ ...$ \\
 & & \vspace{-0.40cm} \\
\hline
 & & \vspace{-0.40cm} \\
$\gamma_x \hat{n}_x - \gamma_y \hat{n}_y$ (and cyclic) & $E$ & $3/2 \ , \ 5/2 \ , \ ...$ \\
 & & \vspace{-0.40cm} \\
$\gamma_5 (\gamma_x \hat{n}_x - \gamma_y \hat{n}_y)$ (and cyclic) & & $3/2 \ , \ 5/2 \ , \ ...$\vspace{-0.40cm} \\
 & & \\
\hline
\end{tabular}

\caption{\label{TAB002}static-light meson creation operators.}
\end{center}
\end{table}

To optimize the ground state overlap of these static-light meson states, we use Gaussian smearing \cite{Gusken:1989qx} for light quark operators and APE smearing \cite{Albanese:1987ds} for spatial links (parameters $\kappa_\textrm{Gauss} = 0.5$, $N_\textrm{Gauss} = 30$, $\alpha_\textrm{APE} = 0.5$, $N_\textrm{APE} = 10$ and $r = 3$ as in \cite{Jansen:2008si}).


\subsection{Static-light meson masses}

Since we work in the twisted basis, where each of the operators listed in Table~\ref{TAB002} creates both $\mathcal{P} = +$ and $\mathcal{P} = -$ states, it is convenient to determine $\mathcal{P} = +$ and $\mathcal{P} = -$ static-light meson masses from the same correlation matrix.

\begin{figure}[h!]
\begin{center}
\input{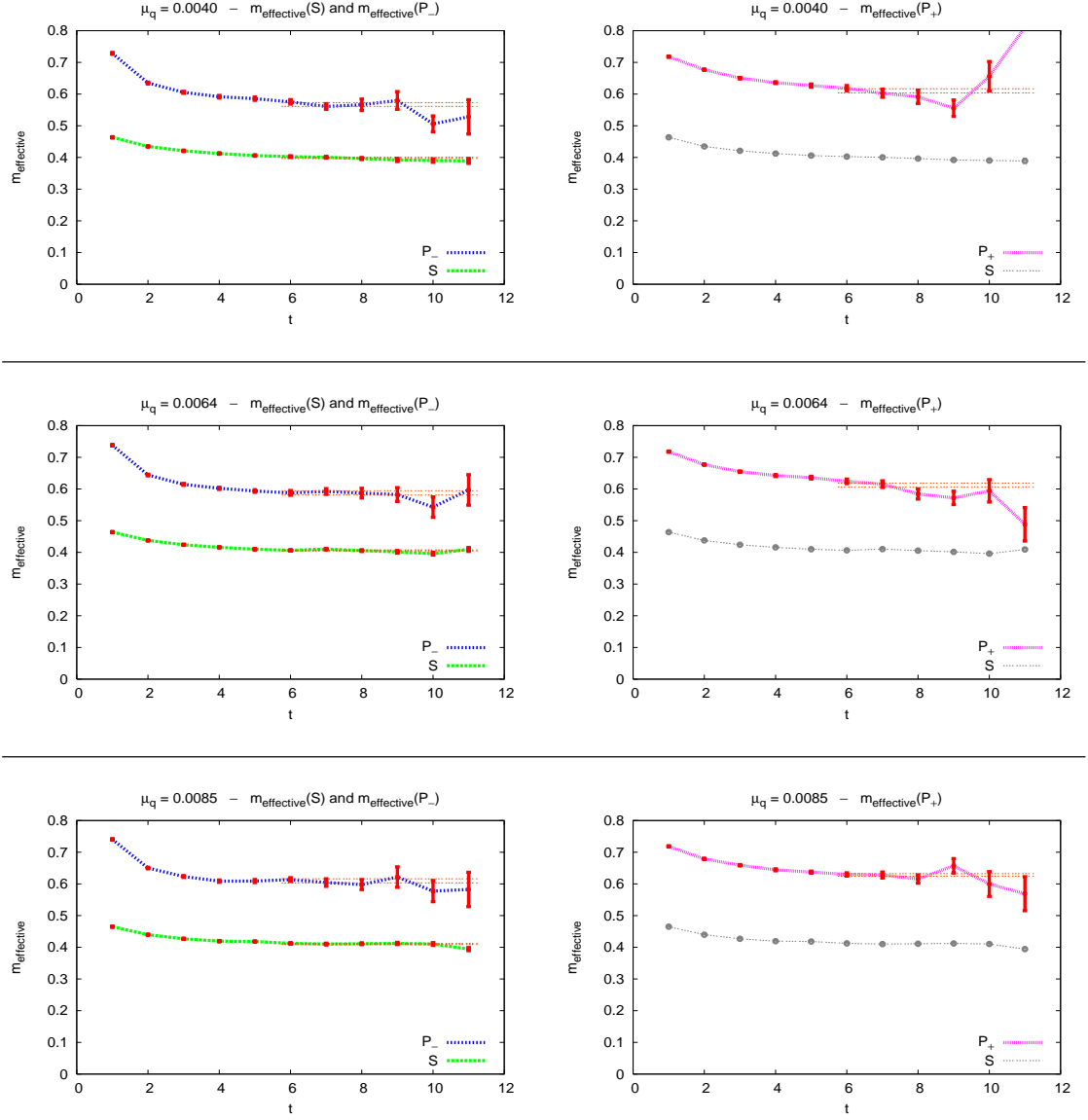}
\caption{\label{FIG001}effective masses for $S$, $P_-$ and $P_+$ for $\mu_\textrm{q} \in \{ 0.0040 \, , \, 0.0064 \, , \, 0.0085 \}$.}
\end{center}
\end{figure}

For $S$ and $P_-$ we compute the $2 \times 2$ matrix
\begin{eqnarray}
\mathcal{C}_{J K}(t) \ \ = \ \ \Big\langle \Big(\mathcal{O}^{(\Gamma_J)}(t)\Big)^\dagger \mathcal{O}^{(\Gamma_K)}(0) \Big\rangle ,
\end{eqnarray}
where $\Gamma_J \in \{ \gamma_5 \, , \, 1 \}$, and solve the generalized eigenvalue problem
\begin{eqnarray}
\label{EQN691} \mathcal{C}_{J K}(t) v_K^{(n)}(t) \ \ = \ \ \mathcal{C}_{J K}(t_0) v_K^{(n)}(t) 
\lambda^{(n)}(t, t_0) \quad , \quad t_0 \ \ = \ \ 1
\end{eqnarray}
(cf.\ \cite{Luscher:1990ck,Blossier:2009kd}). The meson masses $m(S)$ and $m(P_-)$ are determined by performing $\chi^2$ minimizing fits to effective mass plateaus,
\begin{eqnarray}
m_\textrm{effective}^{(n)}(t) \ \ = \ \ \ln\bigg(\frac{\lambda^{(n)}(t,t_0)}{\lambda^{(n)}(t+1,t_0)}\bigg) ,
\end{eqnarray}
at large temporal separations $t$ (as indicated in Figure~\ref{FIG001} our fitting range is $6 \leq t \leq 11$). The parity of the corresponding states, i.e.\ whether it is $S$ or $P_-$, can be extracted from the eigenvectors $v_J^{(n)}$ (for a detailed discussion, of how to identify parity, cf.\ \cite{Jansen:2008si}). Results of meson masses and mass differences and corresponding reduced $\chi^2$ values are listed in Table~\ref{TAB034}.

\begin{table}[h!]
\begin{center}

\begin{tabular}{|c||c|c||c|c||c|c|}
\hline
 & & & & & & \vspace{-0.40cm} \\
$\mu_\textrm{q}$ & $m(S)$ & $\chi^2 / \textrm{dof}$ & $m(P_-)$ & $\chi^2 / \textrm{dof}$ & $m(P_+)$ & $\chi^2 / \textrm{dof}$ \\
 & & & & & & \vspace{-0.40cm} \\
\hline
 & & & & & & \vspace{-0.40cm} \\
$0.0040$ & $0.3987(19)$ & $1.79$ & $0.5670(60)$ & $1.69$ & $0.6101(66)$ & $2.46$ \\
$0.0064$ & $0.4061(17)$ & $1.93$ & $0.5877(67)$ & $0.45$ & $0.6121(64)$ & $3.01$ \\
$0.0085$ & $0.4104(17)$ & $2.23$ & $0.6095(65)$ & $0.49$ & $0.6283(41)$ & $0.87$\vspace{-0.40cm} \\
 & & & & & & \\
\hline
\end{tabular}

\vspace{0.3cm}

\begin{tabular}{|c||c|c|}
\hline
 & & \vspace{-0.40cm} \\
$\mu_\textrm{q}$ & $m(P_-) - m(S)$ & $m(P_+) - m(S)$ \\
 & & \vspace{-0.40cm} \\
\hline
 & & \vspace{-0.40cm} \\
$0.0040$ & $0.1683(65)$ & $0.2114(62)$ \\
$0.0064$ & $0.1817(69)$ & $0.2060(63)$ \\
$0.0085$ & $0.1991(63)$ & $0.2179(41)$\vspace{-0.40cm} \\
 & & \\
\hline
\end{tabular}










\caption{\label{TAB034}static-light meson masses and mass differences for $\mu_\textrm{q} \in \{ 0.0040 \, , \, 0.0064 \, , \, 0.0085 \}$.}
\end{center}
\end{table}

For $m(P_+)$ we proceed analogously this time computing the $2 \times 2$ matrix (\ref{EQN691}), where \\ $\Gamma_J \in \{ \gamma_x \hat{n}_x - \gamma_y \hat{n}_y \, , \, \gamma_5 (\gamma_x \hat{n}_x - \gamma_y \hat{n}_y) \}$.

By solving the generalized eigenvalue problem (\ref{EQN691}) we have also obtained appropriate linear combinations of 
twisted basis meson creation operators with well defined parity. To be more precise the operators
\begin{eqnarray}
\label{EQN696} & & \hspace{-0.7cm} \mathcal{O}^{(S)} \ \ = \ \ v_{\gamma_5}^{(S)}(t) \mathcal{O}^{(\gamma_5)} + v_{1}^{(S)}(t) \mathcal{O}^{(1)} \\
 & & \hspace{-0.7cm} \mathcal{O}^{(P_-)} \ \ = \ \ v_{\gamma_5}^{(P_-)}(t) \mathcal{O}^{(\gamma_5)} + v_{1}^{(P_-)}(t) \mathcal{O}^{(1)} \\
\label{EQN698} & & \hspace{-0.7cm} \mathcal{O}^{(P_+)} \ \ = \ \ v_{\gamma_x \hat{n}_x - \gamma_y \hat{n}_y}^{(P_+)}(t) \mathcal{O}^{(\gamma_x \hat{n}_x - \gamma_y \hat{n}_y)} + v_{\gamma_5 (\gamma_x \hat{n}_x - \gamma_y \hat{n}_y)}^{(P_+)}(t) \mathcal{O}^{(\gamma_5 (\gamma_x \hat{n}_x - \gamma_y \hat{n}_y))}
\end{eqnarray}
create static-light meson states, which have the same quantum numbers $j^\mathcal{P}$ as the states of interest, 
$| S \rangle$, $| P_- \rangle$ and $| P_+ \rangle$ respectively. Since the $t$ dependence of the eigenvectors 
$v_J^{(n)}$ is very weak \cite{Gattringer:2007da}, results are essentially unaffected by the choice of $t$ 
(we have used $t = 6$ for all results presented in the following).


\subsection{Two-point functions and their ground state norms}

After having obtained the linear combinations of twisted basis operators (\ref{EQN696}) to (\ref{EQN698}) the two-point functions
\begin{eqnarray}
\Big\langle \Big(\mathcal{O}^{(S)}(t)\Big)^\dagger \mathcal{O}^{(S)}(0) \Big\rangle \quad , \quad \Big\langle \Big(\mathcal{O}^{(P_-)}(t)\Big)^\dagger \mathcal{O}^{(P_-)}(0) \Big\rangle \quad , \quad \Big\langle \Big(\mathcal{O}^{(P_+)}(t)\Big)^\dagger \mathcal{O}^{(P_+)}(0) \Big\rangle
\end{eqnarray}
are straightforward to compute.

From these two-point functions we also determine the ground state norms of the corresponding 
$j^\mathcal{P}$ sectors, $N(S)$, $N(P_-)$ and $N(P_+)$, 
by fitting exponentials at large temporal separations. To be more precise, we obtain e.g.\ $N(S)$ by fitting 
$N(S)^2 e^{-m t}$ to $\langle (\mathcal{O}^{(S)}(t))^\dagger \mathcal{O}^{(S)}(0) \rangle$ with $N(S)$ and $m$ 
as degrees of freedom. Results and corresponding reduced $\chi^2$ values are listed in 
Table~\ref{TAB692} (fitting range $6 \leq t \leq 12$).

\begin{table}[h!]
\begin{center}

\begin{tabular}{|c||c|c||c|c||c|c|}
\hline
 & & & & & & \vspace{-0.40cm} \\
$\mu_\textrm{q}$ & $N(S)$ & $\chi^2 / \textrm{dof}$ & $N(P_-)$ & $\chi^2 / \textrm{dof}$ & $N(P_+)$ & $\chi^2 / \textrm{dof}$ \\
 & & & & & & \vspace{-0.40cm} \\
\hline
 & & & & & & \vspace{-0.40cm} \\
$0.0040$ & $0.3271(26)$ & $0.21$ & $0.2998(93) \ \, $ & $0.33$ & $0.1139(26)$ & $1.43$ \\
$0.0064$ & $0.3358(20)$ & $0.23$ & $0.3074(87) \ \, $ & $0.13$ & $0.1120(27)$ & $1.68$ \\
$0.0085$ & $0.3397(22)$ & $0.22$ & $0.3139(103)$ & $0.08$ & $0.1212(22)$ & $0.28$\vspace{-0.40cm} \\
 & & & & & & \\
\hline
\end{tabular}







\caption{\label{TAB692}ground state norms for $\mu_\textrm{q} \in \{ 0.0040 \, , \, 0.0064 \, , \, 0.0085 \}$.}
\end{center}
\end{table}


\subsection{Three-point functions and form factors $\tau_{1/2}$ and $\tau_{3/2}$}

In analogy to effective masses we define effective form factors
\begin{eqnarray}
\nonumber & & \hspace{-0.7cm} \tau_{1/2 , \textrm{effective}}(t_0-t_1,t_1-t_2) \
\  \\
\label{EQN795} & & = \ \ \frac{1}{Z_\mathcal{D}} \Bigg|\frac{N(P_-) \ \ N(S) \ \
\Big\langle \Big(\mathcal{O}^{(P_-)}(t_0)\Big)^\dagger \ (\bar{Q} \gamma_5
\gamma_z D_z Q)(t_1) \ \mathcal{O}^{(S)}(t_2) \Big\rangle}{\Big(m(P_-) -
m(S)\Big) \ \ \Big\langle \Big(\mathcal{O}^{(P_-)}(t_0)\Big)^\dagger
\mathcal{O}^{(P_-)}(t_1) \Big\rangle \ \ \Big\langle
\Big(\mathcal{O}^{(S)}(t_1)\Big)^\dagger \mathcal{O}^{(S)}(t_2)
\Big\rangle}\Bigg| \\
\nonumber & & \hspace{-0.7cm} \tau_{3/2 , \textrm{effective}}(t_0-t_1,t_1-t_2) \
\  \\
\label{EQN796} & & = \ \ \frac{1}{Z_\mathcal{D}} \Bigg|\frac{N(P_+) \ \ N(S) \ \
\Big\langle \Big(\mathcal{O}^{(P_+)}(t_0)\Big)^\dagger \ (\bar{Q} \gamma_5
(\gamma_x D_x - \gamma_y D_y) Q)(t_1) \ \mathcal{O}^{(S)}(t_2)
\Big\rangle}{\sqrt{6} \ \ \Big(m(P_+) - m(S)\Big) \ \ \Big\langle
\Big(\mathcal{O}^{(P_+)}(t_0)\Big)^\dagger \mathcal{O}^{(P_+)}(t_1) \Big\rangle
\ \ \Big\langle \Big(\mathcal{O}^{(S)}(t_1)\Big)^\dagger \mathcal{O}^{(S)}(t_2)
\Big\rangle}\Bigg|
\end{eqnarray}
($Z_\mathcal{D} = 0.976$ is a lattice renormalization constant, which we derive and discuss in detail in section~\ref{SEC_renorm}). These effective form factors are related to $\tau_{1/2}$ and $\tau_{3/2}$ via (\ref{EQN631}) and (\ref{EQN632}):
\begin{eqnarray}
 & & \hspace{-0.7cm} \tau_{1/2}(1) \ \ = \ \ \lim_{t_0-t_1 \rightarrow \infty \, , \, t_1-t_2 \rightarrow \infty} \tau_{1/2 , \textrm{effective}}(t_0-t_1,t_1-t_2) \\
 & & \hspace{-0.7cm} \tau_{3/2}(1) \ \ = \ \ \lim_{t_0-t_1 \rightarrow \infty \, , \, t_1-t_2 \rightarrow \infty} \tau_{3/2 , \textrm{effective}}(t_0-t_1,t_1-t_2) .
\end{eqnarray}

Computation of the three-point functions appearing in (\ref{EQN795}) and (\ref{EQN796}) is again straightforward. 
We chose to represent the covariant derivative acting on the static quark field symmetrically by
\begin{eqnarray}
D_j Q(\mathbf{x},t) \ \ = \ \ \frac{1}{2} \Big(U_j(\mathbf{x},t) Q(\mathbf{x}+\mathbf{e}_j,t) - \Big(U_j(\mathbf{x}-\mathbf{e}_j,t)\Big)^\dagger Q(\mathbf{x}-\mathbf{e}_j,t)\Big) .
\end{eqnarray}
To optimally exploit our gauge configurations and propagator inversions, we average over all three-point functions, 
which are related by the lattice symmetries $\gamma_5$ hermiticity, parity, time reversal, charge conjugation and cubic 
rotations.

\begin{figure}[b!]
\begin{center}
\input{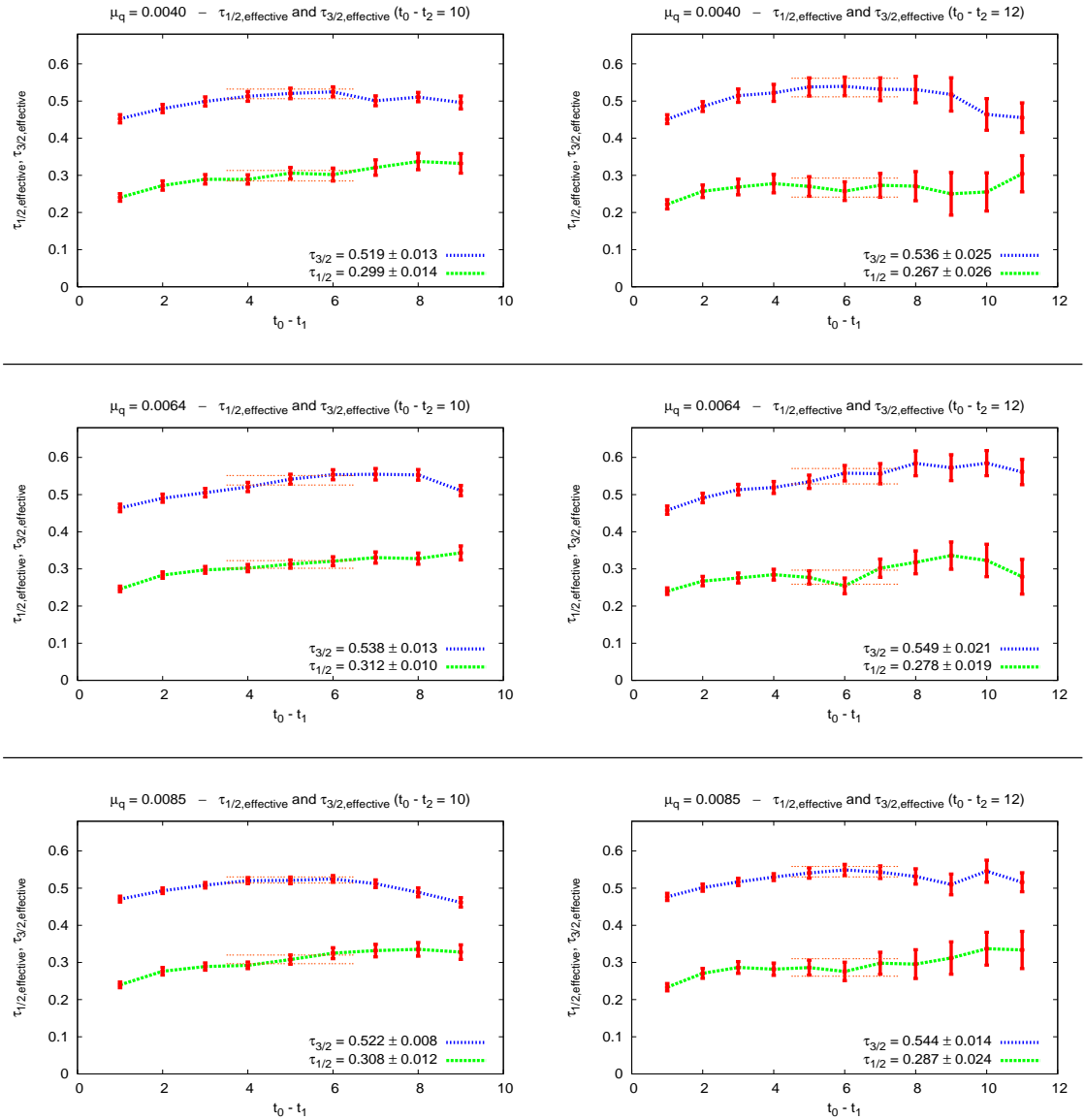}
\caption{\label{FIG002}effective form factors $\tau_{1/2 , \textrm{effective}}$
and $\tau_{3/2 , \textrm{effective}}$ for $t_0 - t_2 \in \{ 10 \, , \, 12 \}$
and $\quad \quad \quad \quad \quad$ $\mu_\textrm{q} \in \{ 0.0040 \, , \, 0.0064
\, , \, 0.0085 \}$.}
\end{center}
\end{figure}

The resulting effective form factors $\tau_{1/2 , \textrm{effective}}(t_0-t_1,t_1-t_2)$ and 
$\tau_{3/2 , \textrm{effective}}(t_0-t_1,t_1-t_2)$ are shown in Figure~\ref{FIG002} as functions of 
$t_0 - t_1$ for fixed $t_0 - t_2 \in \{ 10 \, , \, 12 \}$. Within statistical errors these effective form factors 
exhibit plateaus for $t_0 - t_1 \approx (t_0 - t_2) / 2$, i.e.\ when both temporal separations, $t_0 - t_1$ and 
$t_1 - t_2$, are large. We determine $\tau_{1/2}$ and $\tau_{3/2}$ by performing $\chi^2$ minimizing fits to the 
central three data points as indicated in Figure~\ref{FIG002}. Results for $t_0 - t_2 = 10$ and for 
$t_0 - t_2 = 12$, which are listed in Table~\ref{TAB438}, are in agreement within statistical errors. We consider 
this a strong indication that contributions from excited states at these temporal separations are essentially 
negligible and that the plateaus of the effective form factors indeed correspond to $\tau_{1/2}$ and $\tau_{3/2}$. 
In the following discussions we only quote the numbers corresponding to $t_0 - t_2 = 10$, since their statistical 
errors are significantly smaller than those for $t_0 - t_2 = 12$. 

\begin{table}[h!]
\begin{center}

\begin{tabular}{|c||c||c|c|c|c|}
\hline
 & & & & & \vspace{-0.40cm} \\
$\mu_\textrm{q}$ & $t_0 - t_2$ & $\tau_{1/2}$ & $\tau_{3/2}$ & $\tau_{3/2} / \tau_{1/2}$ & $(\tau_{3/2})^2 - (\tau_{1/2})^2$ \\
 & & & & & \vspace{-0.40cm} \\
\hline
 & & & & & \vspace{-0.40cm} \\
$0.0040$ & $10$ & $0.299(14)$ & $0.519(13)$ & $1.74(9) \ \, $ & $0.180(16)$ \\
         & $12$ & $0.267(26)$ & $0.536(25)$ & $2.01(21)$ & $0.216(30)$ \\
 & & & & & \vspace{-0.40cm} \\
\hline
 & & & & & \vspace{-0.40cm} \\
$0.0064$ & $10$ & $0.312(10)$ & $0.538(13)$ & $1.73(6) \ \, $ & $0.193(13)$ \\
         & $12$ & $0.278(19)$ & $0.549(21)$ & $1.98(14)$ & $0.225(23)$ \\
 & & & & & \vspace{-0.40cm} \\
\hline
 & & & & & \vspace{-0.40cm} \\
$0.0085$ & $10$ & $0.308(12)$ & $0.522(8) \ \, $ & $1.69(6) \ \, $ & $0.177(9) \ \, $ \\
         & $12$ & $0.287(24)$ & $0.544(14)$ & $1.90(17)$ & $0.214(21)$\vspace{-0.40cm} \\
 & & & & & \\
\hline
\end{tabular}






















\caption{\label{TAB438}$\tau_{1/2}$ and $\tau_{3/2}$ for $t_0 - t_2 \in \{ 10 \,
, \, 12 \}$ and $\mu_\textrm{q} \in \{ 0.0040 \, , \, 0.0064 \, , \, 0.0085
\}$.}
\end{center}
\end{table}

As expected from operator product expansion, $\tau_{3/2}(1)$ is significantly
larger than $\tau_{1/2}(1)$.  Moreover the Uraltsev sum rule
\cite{Uraltsev:2000ce},
\begin{eqnarray}
\sum_n \Big|\tau_{3/2}^{(n)}(1)\Big|^2 - \Big|\tau_{1/2}^{(n)}(1)\Big|^2 \ \ = \ \ \frac{1}{4} ,
\end{eqnarray}
is almost fulfilled by the ground state contributions $\tau_{1/2}^{(0)}(1)
\equiv \tau_{1/2}(1)$ and  $\tau_{3/2}^{(0)}(1) \equiv \tau_{3/2}(1)$.

\begin{figure}[h!]
\begin{center}
\input{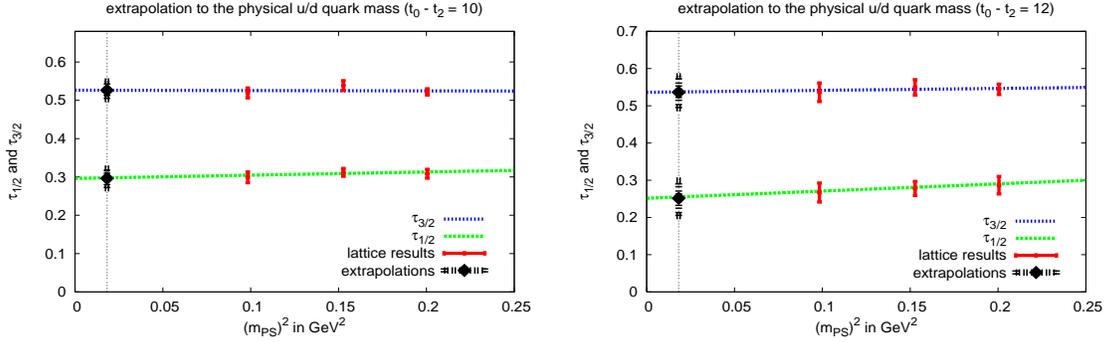}
\caption{\label{FIG003}linear extrapolation of $\tau_{1/2}$ and $\tau_{3/2}$ to the $u/d$ quark mass for $t_0 - t_2 \in \{ 10 \, , \, 12 \}$.}
\end{center}
\end{figure}

Finally we use our results at three different light quark masses (cf.\ Table~\ref{TAB001}) to perform 
a linear extrapolation of the form factors in $(m_\textrm{PS})^2$ to the physical $u/d$ quark mass ($m_\textrm{PS} = 135 \, \textrm{MeV}$). Results are shown in Figure~\ref{FIG003} and 
Table~\ref{TAB237}. The qualitative picture for $u/d$ quark masses is the same as for the heavier masses used 
directly in our simulations: $\tau^{m_{\rm phys}}_{3/2}(1) = 0.526(23)$ is significantly larger than 
$\tau^{m_{\rm phys}}_{1/2}(1) = 0.296(26)$ 
supporting the ``theory expectation'' that a decay of a $B$ meson to a $j = 3/2$ $P$ wave $D$ meson is more likely than to a $j = 1/2$ $P$ wave $D$ meson.

\begin{table}[h!]
\begin{center}

\begin{tabular}{|c||c|c||c|c|}
\hline
 & & & & \vspace{-0.40cm} \\
$t_0 - t_2$ & $\tau_{1/2}(1)$ & $\chi^2 / \textrm{dof}$ & $\tau_{3/2}(1)$ & $\chi^2 / \textrm{dof}$ \\
 & & & & \vspace{-0.40cm} \\
\hline
 & & & & \vspace{-0.40cm} \\
$10$ & $0.296(26)$ & $0.34$ & $0.526(23)$ & $1.43$ \\
$12$ & $0.251(48)$ & $0.00$ & $0.536(43)$ & $0.12$\vspace{-0.4cm} \\
 & & & & \\
\hline
\end{tabular}\label{table:final}








\caption{\label{TAB237}linear extrapolation of $\tau_{1/2}$ and $\tau_{3/2}$ to the $u/d$ quark mass for $t_0 - t_2 \in \{ 10 \, , \, 12 \}$.}
\end{center}
\end{table}



\section{\label{SEC_renorm}Perturbative renormalization of the static current
$\bar{Q} \gamma_5 \gamma_i D_j Q$}

In this section we derive the analytical formulae and give the numerical values
of the renormalization constant $Z_\mathcal{D}$ of the dimension 4 current $O_{i
j} = \bar{Q} \gamma_5 \gamma_i D_j Q$ computed at first order of perturbation
theory for the HYP smeared static quark action and both the standard Wilson
plaquette and the tree-level Symanzik improved gauge action.


\subsection{Definitions}

The bare propagator of a static quark on the lattice is
\begin{eqnarray}
\nonumber & & \hspace{-0.7cm} S^B(p) \ \ = \ \ \frac{a}{1 - e^{-i p_4 a} + a \delta m + a \Sigma(p)} \ \ = \ \ \frac{a}{1 - e^{-i p_4 a}} \sum_n \left(-\frac{a (\delta m + \Sigma(p))}{1 - e^{-i p_4 a}}\right)^n \ \ \\
\label{propag} & & \equiv \ \ Z_{2 h} S^R(p) .
\end{eqnarray}
Choosing the renormalization conditions 
\begin{eqnarray}
(S^R)^{-1}(p)\Big|_{i p_4 \to 0} \ \ = \ \ i p_4 \quad , \quad \delta m \ \ = \ \ -\Sigma(p_4=0)
\end{eqnarray}
implies
\begin{eqnarray}
Z_{2 h} \ \ = \ \ 1 - \frac{d \Sigma}{d (i p_4)}\bigg|_{i p_4 \to 0} .
\end{eqnarray}
The bare vertex function $V^B_{i j}(p)$ is defined as
\begin{eqnarray}
\nonumber & & \hspace{-0.7cm} V^B_{i j}(p) \ \ = \ \ (S^B)^{-1}(p) \sum_{x,y} e^{i p (x-y)} \Big\langle Q^B(x) O^B_{i j}(0) \bar{Q}^B(y) \Big\rangle (S^B)^{-1}(p) \ \  \\
\label{vertex} & & = \ \ \frac{Z_\mathcal{D}}{Z_{2 h}} (S^R)^{-1}(p) \sum_{x,y} e^{i p (x-y)} \Big\langle Q^R(x) O^R_{i j}(0) \bar{Q}^R(y) \Big\rangle (S^R)^{-1}(p) ,
\end{eqnarray}
where
\begin{eqnarray}
O^B_{i j}(0) \ \ = \ \ Z_\mathcal{D} O^R_{i j}(0) .
\end{eqnarray}

$V^B_{i j}(p)$ can be written as
\begin{eqnarray}
V^B_{i j}(p) \ \ = \ \ (1 + \delta V) \bar{u}(p) \gamma_i \gamma_5 p_j u(p) \ \ \equiv \ \ (1 + \delta V) V^R_{i j}(p) .
\end{eqnarray}
$\delta V$ is given by all the 1PI one-loop diagrams containing the vertex.


\subsection{Analytical formulae and results}

The notations used in this section and the Feynman rules are listed in appendix~\ref{sec2}. They are the same as in \cite{Blossier:2005vy} except for the gluon propagator having the form
\begin{eqnarray}
\label{gluonpropagator} D_{\mu \nu} \ \ = \ \ C^{-1}_0 D^\textrm{plaq}_{\mu \nu}
+ \Delta_{\mu \nu}
\end{eqnarray}
\cite{Horsley:2004mx}, where $C_0 = c_0 + 8 c_1 + 16 c_2 + 8 c_3 \equiv 1$, $c_1
= -1/12$, $c_2 = c_3 = 0$ for the case of the tree-level Symanzik improved gauge
action and
\begin{eqnarray}
\Delta_{\mu \nu} \ \ = \ \ \delta_{\mu \nu} K_\mu + 4 L_{\mu \nu} N_\mu N_\nu .
\end{eqnarray}
Finally $K_\mu$ and $L_{\mu \nu}$ are complicated expressions, which do not need
to be reproduced here. The only relevant features for this work are that
$\Delta_{\mu \nu}$ is regular in the infrared regime and $K_\mu = K^0 + 4
N^2_\mu K'_\mu$.

\begin{figure}[b]

\begin{center}

\begin{tabular}{ccc}

\begin{picture}(-10,30)(10,30)

\Line(-45,30)(55,30)
\Line(-45,31)(55,31)
\CTri(-23,27)(-23,34)(-18,30.5){Black}{Black}
\CTri(2,27)(2,34)(7,30.5){Black}{Black}
\CTri(27,27)(27,34)(32,30.5){Black}{Black}
\GlueArc(5,30.5)(12.5,0,180){1}{10}
\Text(30,20)[c]{\small{$p$}}
\Text(5,20)[c]{\small{$p$+$k$}}
\Text(-20,20)[c]{\small{$p$}}

\end{picture}

&\rule[0cm]{2.0cm}{0cm}&
\begin{picture}(-10,30)(10,30)

\Line(-35,30)(45,30)
\Line(-35,31)(45,31)
\CTri(-18,27)(-18,34)(-13,30.5){Black}{Black}
\CTri(22,27)(22,34)(27,30.5){Black}{Black}
\GlueArc(5,44)(12.5,0,360){1}{20}
\Text(25,20)[c]{\small{$p$}}
\Text(-15,20)[c]{\small{$p$}}

\end{picture}
\\
\\
(a): sunset diagram&\rule[0cm]{2.0cm}{0cm}&
(b): tadpole diagram\\
\end{tabular}
\end{center}
\caption{\label{self}self-energy corrections.}
\end{figure}
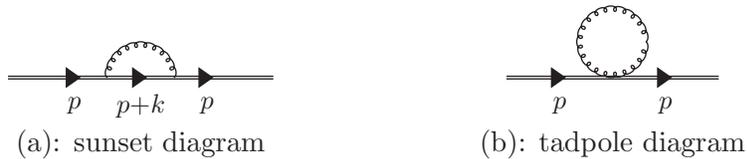

The static quark self-energy expressed at the first order of perturbation theory
is given by $\Sigma(p) = -(F_1+F_2)$, where $F_1$ and $F_2$ correspond to the
diagrams shown in Figure~\ref{self}(a) and (b): %
\begin{eqnarray}
 & & \hspace{-0.7cm} F_1 \ \ = \ \ -\frac{4}{3 a} g^2_0 \int_k h_{4 i} h_{4 j}
D_{i j} \frac{e^{-i (k_4 + 2 a p_4)}}{1 - e^{-i (k_4 + a p_4)} + \epsilon} \ \ =
\ \ F^\textrm{plaq}_1 + F'_1
\end{eqnarray}
\begin{eqnarray}
\nonumber & & \hspace{-0.7cm} F^\textrm{plaq}_1 \ \ = \ \ -\frac{4}{3 a} g^2_0
\int_k \frac{D^2_4 + \sum_{i=1}^3 G_{4 i}^2}{2 W + a^2 \lambda^2} \frac{e^{-i
(k_4 + 2 a p_4)}}{1 - e^{-i (k_4 + a p_4)}+ \epsilon} \ \  \\
\nonumber & & =_{a p_4 \to 0} \ \ \frac{4}{3 a} g^2_0 \int_{\vec{k}} \frac{D^2_4
(-iE) + \sum_{i=1}^3 G^2_{4 i}(-i E)}{4 E \sqrt{1 + E^2}}\frac{1}{1 - e^{E'}} \\
\label{diag1} & & \hspace{0.675cm} + \frac{4}{3} g^2_0 i p_4 \int_{\vec{k}} \frac{D^2_4 (-i E) + \sum_{i=1}^3 G^2_{4
i}(-i E)}{2 E \sqrt{1 + E^2}} \bigg[\frac{1}{e^{E'} - 1} + \frac{1}{2} \frac{1}{(e^{E'} - 1)^2}\bigg]
\end{eqnarray}
\begin{eqnarray}
\nonumber & & \hspace{-0.7cm} F'_1 \ \ = \ \ -\frac{4}{3 a} g^2_0 \int_k h_{4 i}
h_{4 j} \Delta_{i j} \frac{e^{-i (k_4 + 2 a p_4)}}{1 - e^{-i (k_4 + a p_4)} +
\epsilon} \ \ = \\
\nonumber & & =_{a p_4 \to 0} \ \ -\frac{4}{3 a} g^2_0 \int_k \frac{M_4 - i
N_4}{2 i N_4 + \epsilon M_4} \Big(D^2_4 K^0 + N^2_4 \Lambda\Big) \\
\nonumber & & \hspace{0.675cm} + \frac{8}{3} g^2_0 i p_4 \int_k \bigg[\frac{M_4
- i N_4}{2 i N_4 + \epsilon M_4} +\frac{1}{2} \bigg(\frac{M_4 - i N_4}{2 i N_4 +
\epsilon M_4}\bigg)^2\bigg] \Big(D^2_4 K^0 + N^2_4 \Lambda\Big) \ \  \\
 & & = \ \ \frac{2}{3 a} g^2_0 \int_k \Big(D^2_4 K^0 + N^2_4 \Lambda\Big) -
 \frac{1}{3} g^2_0 i p_4 \int_k \Big[M^2_4 \Lambda + 3 \Big(D^2_4 K^0 + N^2_4
 \Lambda\Big)\Big]
\end{eqnarray}
\begin{eqnarray}\nonumber
N^2_4 \Lambda & = & 4 \bigg(D^2_4 N^2_4 (K'_4 + L_{4 4}) + \frac{1}{4} 
\sum_i G^2_{4i} (K^0 + 4N^2_i K'_i)+ 2 D_4 N_4
\sum_{i=1}^3 G_{4 i} N_i L_{4 i} \\
&&\hspace{0.5cm} + 2 \sum_{i,j=1}^3 G_{4 i} G_{4 j} N_i N_j L_{i
j}\bigg)
\end{eqnarray}
\begin{eqnarray}
\label{f2} F_2 \ \ = \ \ -\frac{1}{2} \frac{4 g^2_0}{3 a} e^{-i a p_4} \int_k
h_{4 i} h_{4 j}D_{i j} \ \ = \ \ F^\textrm{plaq}_2 + F'_2
\end{eqnarray}
\begin{eqnarray}
\nonumber & & \hspace{-0.7cm} F^\textrm{plaq}_2 \ \ = \ \ -\frac{1}{2} \frac{4 g^2_0}{3 a} e^{-i a p_4} \int_k \frac{D^2_4 + \sum_{i=1}^3 G^2_{4 i}}{2 W} \ \  \\
\label{f2plaq} & & =_{a p_4 \to 0} \ \ -\frac{1}{2} \frac{4 g^2_0}{3} \Big(1/a - i p_4\Big) \int_k \frac{D^2_4 + \sum_{i=1}^3 G^2_{4 i}}{2 W}
\end{eqnarray}
\begin{eqnarray}
F'_2 \ \ = \ \ -\frac{1}{2} \frac{4 g^2_0}{3 a} e^{-i a p_4} \int_k h_{4 i} h_{4 j} \Delta_{i j} \ \ =_{a p_4 \to 0} \ \ -\frac{1}{2} \frac{4 g^2_0}{3} \Big(1/a - i p_4\Big) \int_k (D^2_4 K^0 + N^2_4 \Lambda) .
\end{eqnarray}
The factor $1/2$ has been introduced to compensate the over-counting of the
factor $2$ in the Feynman rule of the two-gluon vertex, when a closed gluonic
loop is computed.

The other terms entering the above integrals cancel, because the contour can be
closed in the complex plane without including the pole $k_4 = -p_4 + i
\ln(1+\epsilon)$. Finally we can write
\begin{eqnarray}
F_1 \ \ \equiv \ \ -\frac{g^2_0}{12 \pi^2} 
\bigg[\Big(f^\textrm{plaq}_1(\alpha_i) + f'_1(\alpha_i,c_i)\Big) / a + i p_4
\Big(2 \ln(a^2 \lambda^2)  + f^\textrm{plaq}_2(\alpha_i) +
f'_2(\alpha_i,c_i)\Big)\bigg]
\end{eqnarray}
\begin{eqnarray}
F_2 \ \ \equiv \ \ -\frac{g^2_0}{12 \pi^2} \Big(1/a - i p_4\Big)
\Big(f^\textrm{plaq}_3(\alpha_i) + f'_3(\alpha_i,c_i)\Big) .
\end{eqnarray}
The linearly divergent part in $1/a$ of the self-energy is given by
\begin{eqnarray}
\label{selfenergie} \Sigma_0(\alpha_i) \ \ = \ \ \frac{g^2_0}{12 \pi^2 a}
\sigma_0(\alpha_i) \quad , \quad \sigma_0 \ \ = \ \ f_1 + f'_1 + f_3 + f'_3 ,
\end{eqnarray}
while the wave function renormalization $Z_{2 h}$ reads
\begin{eqnarray}
\label{Z_h} Z_{2 h}(\alpha_i) \ \ = \ \ 1 + \frac{g^2_0}{12 \pi^2} \Big(-2 \ln(a^2 \lambda^2) + z_2(\alpha_i)\Big) \quad , \quad z_2 \ \ = \ \ f_3 + f'_3 - (f_2 + f'_2) .
\end{eqnarray}
In Table~\ref{tabzd} we have collected the numerical values of $f_i$, $f'_i$,
$\sigma_0$ and $z_2$ for different kinds of static quark and gluonic actions.

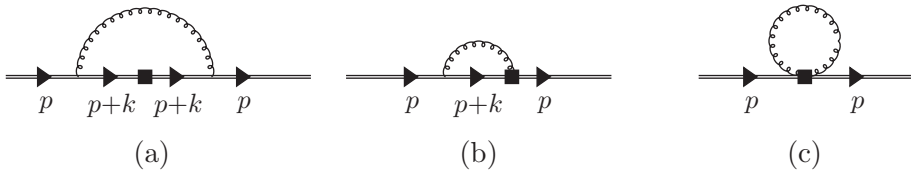
\begin{figure}[b]

\begin{center}

\begin{tabular}{ccccc}

\begin{picture}(-10,10)(10,40)

\Line(-50,30)(65,30)
\Line(-50,31)(65,31)
\CTri(-38,27)(-38,34)(-33,30.5){Black}{Black}
\CTri(-13,27)(-13,34)(-8,30.5){Black}{Black}
\CTri(12,27)(12,34)(17,30.5){Black}{Black}
\CTri(37,27)(37,34)(42,30.5){Black}{Black}
\GlueArc(2.5,30.5)(25,0,180){1}{20}
\Text(40,20)[c]{\small{$p$}}
\Text(15,20)[c]{\small{$p$+$k$}}
\Text(-10,20)[c]{\small{$p$+$k$}}
\Text(-35,20)[c]{\small{$p$}}
\CBox(0,28)(5,33){Black}{Black}
\end{picture}
&\rule[0cm]{3.0cm}{0cm}&

\begin{picture}(-10,10)(10,40)
\Line(-45,30)(55,30)
\Line(-45,31)(55,31)
\CTri(-23,27)(-23,34)(-18,30.5){Black}{Black}
\CTri(2,27)(2,34)(7,30.5){Black}{Black}
\CTri(27,27)(27,34)(32,30.5){Black}{Black}
\GlueArc(5,30.5)(12.5,0,180){1}{10}
\Text(30,20)[c]{\small{$p$}}
\Text(5,20)[c]{\small{$p$+$k$}}
\Text(-20,20)[c]{\small{$p$}}
\CBox(15,28)(20,33){Black}{Black}

\end{picture}

&\rule[0cm]{3.0cm}{0cm}&
\begin{picture}(-10,10)(10,40)

\Line(-35,30)(45,30)
\Line(-35,31)(45,31)
\CTri(-18,27)(-18,34)(-13,30.5){Black}{Black}
\CTri(22,27)(22,34)(27,30.5){Black}{Black}
\GlueArc(5,44)(12.5,0,360){1}{20}
\Text(25,20)[c]{\small{$p$}}
\Text(-15,20)[c]{\small{$p$}}
\CBox(2.5,28)(7.5,33){Black}{Black}

\end{picture}
\\
\vspace{0.5cm}
&&&&\\
(a)&\rule[0cm]{3.0cm}{0cm}&(b)&\rule[0cm]{3.0cm}{0cm}&(c)\\

\end{tabular}
\end{center}
\caption{\label{opder}operator corrections.}
\end{figure}

The vertex function $V^B_{i j}$ is obtained by writing
\begin{eqnarray}
V^B_{i j} \ \ = \ \ V^0_{i j} + V^1_{i j} + V^2_{i j} \quad , \quad V^k_{i j}(\alpha_i) \ \ = \ \ \bar{u}(p) \gamma_i \gamma^5 u(p) V^k_j(\alpha_i) \quad , \quad l = 0,1,2
\end{eqnarray}
corresponding to the diagrams (a), (b) and (c) in Figure~\ref{opder}. The
contribution $V^0_{i j}$ is given by computing
\begin{eqnarray}
V^0_j(\alpha_i) \ \ = \ \ -\frac{4 i}{3 a} g^2_0 \int_k h_{4 k} h_{4 l} D_{k l}
\sin(k + a p)_j \frac{e^{-i (k_4 + 2 a p_4)}}{(1 - e^{-i (k_4 + a p_4)} +
\epsilon)^2} \ \ = \ \ V^{0,\textrm{plaq}}_j + V'^0_j
\end{eqnarray}
\begin{eqnarray}
\nonumber & & \hspace{-0.7cm} V^{0,\textrm{plaq}}_j \ \ = \ \ -\frac{4 i}{3 a} g^2_0 \int_k 
\frac{D^2_4 + \sum_{i=1}^3 G^2_{4 i}}{2 W + a^2 \lambda^2} \sin(k + a p)_j \frac{e^{-i (k_4 + 2 a p_4)}}{(1 - e^{-i (k_4 + a p_4)} + \epsilon)^2} \ \  \\
\nonumber & & = \ \ -\frac{4 i}{3 a} g^2_0 \int_k \frac{D^2_4 + \sum_{i=1}^3 G^2_{4 i}}{2 W + a^2 \lambda^2} \Big(\Gamma_j + a p_j \cos(k_j)\Big) e^{-i a p_4} \bigg(\frac{e^{-i \frac{k_4 + a p_4}{2}}}{1 - e^{-i (k_4 + a p_4)} + \epsilon}\bigg)^2 \ \  \\
\nonumber & & = \ \ -\frac{4 i}{3 a} g^2_0 \int_k \frac{D^2_4 + \sum_{i=1}^3 G^2_{4 i}}{2 W + a^2 \lambda^2} \Big(\Gamma_j + a p_j \cos(k_j)\Big) (1 - i a p_4) \\
\nonumber & & \hspace{0.675cm} \times \frac{1}{\left[2 i \sin\left(\frac{k_4 + a p_4}{2}\right) + e^{i \frac{k_4 + a
p_4}{2}} \epsilon\right]^2} \ \  \\
 & & = \ \ -\frac{4}{3} i g^2_0 p_j \int_k \frac{D^2_4 + \sum_{i=1}^3 G^2_{4 i}}{2 W + a^2 \lambda^2} \frac{\cos(k_j)}{(2 i N_4 + \epsilon M_4)^2}
\end{eqnarray}
\begin{eqnarray}
\nonumber & & \hspace{-0.7cm} V'^0_j \ \ = \ \ -\frac{4 i}{3 a} g^2_0 \int_k h_{4 k} h_{4 l} \Delta_{k l} \Big(\Gamma_j + a p_j \cos(k_j)\Big) e^{-i a p_4} \bigg(\frac{e^{-i \frac{k_4 + a p_4}{2}}}{1 - e^{-i (k_4 + a p_4)} + \epsilon}\bigg)^2 \ \  \\
\nonumber & & = \ \ -\frac{4 i}{3 a} g^2_0 \int_k h_{4 k} h_{4 l} \Delta_{k l} \Big(\Gamma_j + a p_j \cos(k_j)\Big) (1 - i a p_4) \frac{1}{\left(2 i \sin\left(\frac{k_4 + a p_4}{2}\right) + e^{i \frac{k_4 + a p_4}{2}} \epsilon\right)^2} \ \  \\
\nonumber & & = \ \ -\frac{4 i}{3} g^2_0 p_j \int_k \Big(D^2_4 K^0 + N^2_4 \Lambda\Big) \cos(k_j) \frac{1}{(2 i N_4 + \epsilon M_4)^2} \ \ = \ \ \frac{1}{3} g^2_0 i p_j \int_k \Lambda \cos(k_j) .
\end{eqnarray}
The ``sail diagram'' has the following expression:
\begin{eqnarray}
\nonumber V^1_j \ \ = \ \ \frac{4}{3 a} g^2_0 \int_k h_{4 l} D_{l j} \cos\left(\frac{k_j}{2} + a p_j\right) \frac{e^{-i \left(\frac{k_4}{2} + a p_4\right)}}{1 - e^{-i (k_4 + a p_4)} + \epsilon} \ \ = \ \ V^{1,\textrm{plaq}}_j + V'^1_j
\end{eqnarray}
\begin{eqnarray}
\nonumber & & \hspace{-0.7cm} V^{1,\textrm{plaq}}_j \ \ = \ \ \frac{4}{3 a} g^2_0 \int_k \frac{G_{4 j}}{2 W+ a^2 \lambda^2} \Big(M_j - a p_j N_j\Big) \bigg(1-i \frac{a p_4}{2}\bigg) \frac{1}{2 i \sin\left(\frac{k_4 + a p_4}{2}\right) + e^{i \frac{k_4 + a p_4}{2}} \epsilon} \ \  \\
\nonumber & & = \ \ -\frac{4}{3 a} g^2_0 p_j \int_k \frac{G_{4 j} N_j}{2 W + a^2 \lambda^2} \frac{1}{2 i N_4 + \epsilon M_4} \ \ = \ \ \frac{2}{3} g^2_0 i p_j \int_k \frac{G'_{4 j} N_j}{2 W +a^2 \lambda^2} \quad , \\
 & & \hspace{-0.7cm} G_{4 j} \ \ = \ \ N_4 G'_{4 j}
\end{eqnarray}
\begin{eqnarray}
\nonumber & & \hspace{-0.7cm} V'^1_j \ \ = \ \ \frac{4}{3 a} g^2_0 \int_k h_{4 l} \Delta_{l j} \Big(M_j - a p_j N_j\Big) \bigg(1 - i \frac{a p_4}{2}\bigg) \frac{1}{2 i \sin\left(\frac{k_4 + a p_4}{2}\right) + e^{i \frac{k_4 + a p_4}{2}} \epsilon} \ \  \\
\nonumber & & = \ \ -\frac{4}{3} g^2_0 p_j \int_k \Big(4 D_4 N_4 N_j L_{4 j} + N_4 N_j \Lambda'_j\Big) N_j \frac{1}{2 i N_4 +\epsilon M_4} \ \  \\
 & & = \ \ \frac{2}{3} g^2_0 i p_j \int_k N^2_j \Big(4 D_4 L_{4 j} + \Lambda'_j\Big) \quad , \quad N_4 N_j \Lambda'_j \ \ = \ \ \sum_{i=1}^3 G_{4 i} \Delta_{i j} .
\end{eqnarray}
Note that the contribution of the sail diagram to the final result must be doubled, because the gluon leg can be attached to the static line in two different ways. Eventually the tadpole diagram is given by
\begin{eqnarray}
V^2_j(\alpha_i) \ \ = \ \ -\frac{1}{2!} \frac{4}{3} i g^2_0 p_j \int_k D_{4 4} \ \ = \ \ -\frac{i g^2_0}{12 \pi^2} p_j \Big(f_3(\alpha_i=0) + f'_3(\alpha_i=0,c_i)\Big) .
\end{eqnarray}

We finally have
\begin{eqnarray}
\langle H^{**} | O^R_{i j} | H \rangle \ \ = \ \ \frac{1}{Z_\mathcal{D}(\alpha_i)} \langle H^{**} | O^B_{i j} | H \rangle(\alpha_i) ,
\end{eqnarray}
where 
\begin{eqnarray}
\label{ZIW} & & \hspace{-0.7cm} Z_\mathcal{D}(\alpha_i) \ \ = \ \ Z_{2 h}(\alpha_i) \Big(1 + \delta V(\alpha_i)\Big) \\
 & & \hspace{-0.7cm} \delta V(\alpha_i) \ \ \equiv \ \ \frac{g^2_0}{12 \pi^2} \Big(2 \ln(a^2 \lambda^2) + f_4(\alpha_i) + f'_4(\alpha_i,c_i)\Big)
\end{eqnarray}
i.e.\
\begin{eqnarray}
\label{fin} Z_\mathcal{D}(\alpha_i) \ \ = \ \ 1 + \frac{g^2_0}{12 \pi^2} z_d(\alpha_i) \quad , \quad  z_d \ \ = \ \ z_2 + f_4 + f'_4 .
\end{eqnarray}
The numerical values of $z_d$ are collected in Table~\ref{tabzd} for the different kinds of static quark and gluonic actions. With the bare coupling $g^2_0 \equiv 6/\beta$, the tree-level Symanzik improved gauge action at $\beta=3.9$ and the HYP2 static quark action used in our simulations we obtain \\ $Z_\mathcal{D}(\textrm{tlSym},\textrm{HYP2}) = 0.976$.

\begin{table}[h!]
\begin{center}

\begin{tabular}{|c|r|r|r|}
\cline{2-4}
\multicolumn{1}{c|}{} & & & \vspace{-0.40cm} \\
\multicolumn{1}{c|}{}&$\alpha_i=0$&HYP1&HYP2\\
\multicolumn{1}{c|}{} & & & \vspace{-0.40cm} \\
\hline
 & & & \vspace{-0.40cm} \\
$f_1$ & $7.72$ & $1.64$ & $-1.76$ \\
$f'_1(\textrm{tlSym})$ & $2.10$ & $0.14$ & $0.83$ \\
 & & & \vspace{-0.40cm} \\
$f_2$ & $-12.25$ & $1.60$ & $9.58$ \\
$f'_2(\textrm{tlSym})$ & $-3.43$ & $-0.12$ & $-1.50$ \\
 & & & \vspace{-0.40cm} \\
$f_3$ & $12.23$ & $4.12$ & $5.96$ \\
$f'_3(\textrm{tlSym})$ & $-2.10$ & $-0.14$ & $-0.83$ \\
 & & & \vspace{-0.40cm} \\
$f_4$ & $-12.68$ & $-4.95$ & $-0.56$ \\
$f'_4(\textrm{tlSym})$ & $3.56$ & $2.04$ & $1.67 $\\
 & & & \vspace{-0.40cm} \\
\hline
 & & & \vspace{-0.40cm} \\
$\sigma_0$ & $19.95$ & $5.76$ & $4.20$ \\
 & & & \vspace{-0.40cm} \\
$z_2(\textrm{plaq})$ & $24.48$ & $2.52$ & $-3.62$ \\
$z_2(\textrm{tlSym})$ & $25.81$ & $2.50$ & $-2.96$ \\
 & & & \vspace{-0.40cm} \\
$z_d(\textrm{plaq})$ & $11.80$ & $-2.43$ & $-4.19$ \\
$z_d(\textrm{tlSym})$ & $16.69$ & $-0.41$ & $-1.85$\vspace{-0.40cm} \\
 & & & \\
\hline
\end{tabular}

\caption{\label{tabzd}numerical values of the constants $f_1$, $f'_1$, $f_2$, $f'_2$, $f_3$, $f'_3$, $f_4$, $f'_4$, $\sigma_0$, $z_2$ and $z_d$ defined in the text; $\alpha_i = 0$ denotes the unsmeared Eichten-Hill static quark action, while HYP1 and HYP2 are defined in \cite{Hasenfratz:2001hp} and \cite{Della Morte:2005yc} respectively; ``plaq'' denotes the standard Wilson plaquette gauge action, while ``tlSym'' denotes the tree-level Symanzik improved gauge action.}
\end{center}
\end{table}



\section{\label{SEC_concl}Conclusions}

We have computed the form factors $\tau_{1/2}(1)$ and 
$\tau_{3/2}(1)$ in the static limit, which describe (in this limit) the decay $B \to D^{**}$. This decay is presently a puzzle in the sense that sum rules
derived from QCD point towards a dominance of  $\tau_{3/2}(1)$, while
experimental indications point rather in the opposite direction. The aim of this paper has been to check the dominance of $\tau_{3/2}(1)$ in a quantitative way. 

Our final result extrapolated to the physical $u/d$ quark mass is given in Table~\ref{TAB237}. Since we see no systematic dependence on the temporal separation $t_0-t_2$ except for an increase in 
statistical uncertainty, we keep the result at 
$t_0-t_2 = 10$. To the statistical error we add a systematical error of $3 \%$ to account for the 
uncertainty in the computation of the renormalization constant $Z_\mathcal{D}$, which was computed 
perturbatively. We make the ``guesstimate'' of $100 \%$ uncertainty on $1 - Z_\mathcal{D}$, which 
turns out to be very small. Notice that this uncertainty does not apply to the ratio 
$\tau_{3/2}(1) / \tau_{1/2}(1)$ both having the same $Z_\mathcal{D}$ (cf.\ eqns.\ (\ref{EQN795}) and (\ref{EQN796})). We have at this stage no way to estimate systematic uncertainties arising from finite lattice spacing and from finite volume. Therefore, we must consider the errors we quote as incomplete. We end up with 
\begin{eqnarray}
 & & \hspace{-0.7cm} \label{final} \tau_{1/2}(1)\ \ = \ \ 0.296(26) \quad , \quad \tau_{3/2}(1) \ \ = \ \ 0.526(23) \\
 & & \hspace{-0.7cm} \frac{\tau_{3/2}(1)}{\tau_{1/2}(1)} \ \ = \ \ 1.6 \ldots 1.8 \quad , \quad \Big|\tau_{3/2}(1)\Big|^2 - \Big|\tau_{3/2}(1)\Big|^2 \ \ \approx \ \ 0.17 \ldots 0.21
\end{eqnarray}
in fair agreement with the qualitative claim that $\tau_{3/2}$ is significantly larger than $\tau_{1/2}$. Note also that Uraltsev's sum rule is almost saturated by the ground state contributions providing $\approx 80 \%$ of the required $1 / 4$ (cf.\ eqn.\ (\ref{EQN_uraltsev})).

This result does not differ qualitatively from the preliminary quenched 
computation \cite{Becirevic:2004ta}: \\ $\tau_{1/2} = 0.38(5)$ and $\tau_{3/2} = 0.53(8)$. However, we consider the result presented in this paper as standing on a much firmer ground, because it is unquenched, and because the signal is much clearer and more stable thanks to
better analysis procedures. Our 
result (\ref{final}) is also similar to the prediction of a Bakamjian-Thomas 
relativistic quark model~\cite{Morenas:1997nk}, when using a Godfrey-Isgur
interquark potential: $\tau_{1/2} = 0.22$ and $\tau_{3/2} = 0.54$.

Assuming that the heavy quark limit provides reliable indications and that the
standard identification of narrow $D^{**}$ resonances is correct (i.e.\ $D_1(2420)$ ($J=1$) and $D^*_2(2460)$ ($J=2$) correspond to $j = 3/2$ mesons) this points towards the expected dominance of the semileptonic decay of $B$ mesons into these $j=3/2$ states over the decay into $j=1/2$ states.
The latter, labeled as $D^*_0$ ($J=0$) and $D'_1$ ($J=1$) are  identified to some broad structures, which are seen in the semileptonic $B$ decay around similar masses ($2200 \, \textrm{MeV}$ to $2600 \, \textrm{MeV}$).
Remember, however, that the predicted ratio of branching fractions 
$\textrm{Br}(B \to D^{**}_{3/2}) / \textrm{Br}(B \to D^{**}_{1/2})$ is mainly governed by 
$(\tau_{3/2}(1) / \tau_{1/2}(1))^2$ times a rather large ratio of phase-space factors.

It is usually claimed from experiment that the decay into these broad
resonances are not subdominant   as compared to the narrow resonances. A recent
analysis by BABAR \cite{Aubert:2007tra,Aubert:2008ea} finds significant $B\to D^{(*)} \, \pi \, l \, \nu$, but does not give the relative yield of narrow and broad resonances. In a recent paper by BELLE \cite{:2007rb} the four $D^{**}$ states are distinguished. The $B \to D^*_0 \, l \, \nu$ is observed with a comparatively large signal and, assuming the heavy quark limit to be applicable, they fit $\tau_{3/2}(1) = 0.75$ and $\tau_{1/2}(1) = 1.28$. Compared to our result (\ref{final}) this calls for two comments:
\begin{itemize}
 \item[(1)] The $\tau_{3/2}(1)$ shows fair agreement between theory and experiment. This is encouraging, 
 since the narrow resonances are experimentally rather well under control, i.e.\ the narrow resonances 
 are well seen.

 \item[(2)] The experimental $\tau_{1/2}(1)$ is much larger than our prediction. Note, however, that 
 BELLE does not see the other member of the $j=1/2$ doublet, $B \to D'_1 \, l \, \nu$. This is puzzling 
 and the discrepancy concerning $\tau_{1/2}(1)$ should not be taken as final.
\end{itemize}
  
\textit{In view of the impressive convergence of almost all theoretical estimates of $\tau_{1/2}(1)$ and 
$\tau_{3/2}(1)$, in view of our confidence that the result presented in this paper stands on a firm 
ground, we believe that one can consider as established that QCD predicts a clear dominance of the decay 
into $j=3/2$ in the static limit}. 

It still remains to be solved, how to saturate the inclusive semileptonic branching ratio, 
in other words what to add to the $B \to D^{(*)} \, l \, \nu$ and to the narrow $D^{**}$ resonances.
The analyses performed on Class I non-leptonic $B \to D^{**} \pi$ decay do not find any trace of broad structures \cite{Abe:2004cw,Aubert:2006jc}. Invoking factorization, theoretically well under control 
for this kind of process this naturally leads again to $\tau_{1/2}(1)<\tau_{3/2}(1)$.

Experimental work still has to be done. On the theory side, beyond doing the computation at another finer 
lattice spacing to be able to perform a continuum extrapolation (theoretically well
defined, as recalled in Section \ref{SEC_basics}), an estimate of the $1/m_c$ corrections 
would help a lot. To explore that issue a promising method used to study the $B \to D^{(*)} \, l \, \nu$ 
form factors at non-zero recoil \cite{deDivitiisUK, deDivitiisDF} might be helpful. The contributions of 
other states such as negative parity radial excitations should also be considered.
 
Let us conclude by insisting that the issue at clue is of important relevance: any accurate estimate of 
the $V_{cb}$ parameter of the standard model will never be fully convincing as long as the ``$1/2$ versus 
$3/2$ puzzle'' remains unsolved.


\section*{Acknowledgments}

B.B.\ and O.P.\ thank Ikaros Bigi and the other authors of~\cite{Bigi:2007qp} for many discussions on these issues and having stimulated the present work. We also thank Vladimir Galkin, Karl Jansen, Chris Michael, David Palao and Andrea Shindler for many helpful discussions.

This work has been supported in part by the EU Contract No.~MRTN-CT-2006-035482, ``FLAVIAnet'', by the DFG Sonderforschungsbereich/Transregio SFB/TR9-03 and by the project ANR-NT-05-3-43577 (QCDNEXT).

We thank CCIN2P3 in Lyon and the J\"ulich Supercomputing Center (JSC) for having allocated to us computer time, which was used in this work.



\appendix

\section{\label{sec2}Feynman rules}

The lattice HQET action is
\begin{eqnarray}
\label{actionHQET}S^\textrm{HQET} \ \ = \ \ a^3 \sum_n \bigg(Q^\dagger(n) \Big(Q(n) - U^{\dagger,\textrm{HYP}}_4(n - \hat{4}) Q(n - \hat{4})\Big) + a \delta m Q^\dagger(n) Q(n)\bigg) ,
\end{eqnarray}
where $U^\textrm{HYP}_4(n)$ is a link built from hypercubic blocking.

We will use in the rest of this appendix the following notations taken from \cite{Capitani:2002mp,DeGrand:2002va,Lee:2003sk}:
\begin{eqnarray}
 & & \hspace{-0.7cm} \int_p \ \ \equiv \ \ \int_{-\pi/a}^{\pi/a} \frac{d^4p}{(2 \pi)^4} \quad , \quad \int_{\vec{p}} \ \ \equiv \ \ \int_{-\pi/a}^{\pi/a} \frac{d^3p}{(2\pi)^3} \quad , \quad a^4 \sum_n e^{i p n} \ \ = \ \ \delta(p) \\
 & & \hspace{-0.7cm} \int_k \ \ \equiv \ \ \int_{-\pi}^\pi \frac{d^4k}{(2\pi)^4} \quad , \quad \int_{\vec{k}} \ \ \equiv \ \ \int_{-\pi}^\pi \frac{d^3k}{(2\pi)^3} \\
 & & \hspace{-0.7cm} h(n) \ \ = \ \ \int_p e^{i p n} h(p) \\
 & & \hspace{-0.7cm} U_\mu(n) \ \ = \ \ e^{i a g_0 A^a_\mu(n) T^a} \ \ = \ \ 1 + i a g_0 A^a_\mu(n) T^a - \frac{a^2 g_0^2}{2!} A^a_\mu(n) A^b_\mu(n) T^a T^b + \mathcal{O}(g^3_0) \\
 & & \hspace{-0.7cm} U^\textrm{HYP}_\mu(n) \ \ = \ \ e^{i a g_0 B^a_\mu(n) T^a} \ \ = \ \ 1 + i a g_0 B^a_\mu(n) T^a - \frac{a^2 g^2}{2!} B^a_\mu(n) B^b_\mu(n) T^a T^b + \mathcal{O}(g^3_0) \\
 & & \hspace{-0.7cm} A^a_\mu(n) \ \ = \ \ \int_p e^{i p (n + \frac{a}{2})} A^a_\mu(p) \quad , \quad B^a_\mu(n) \ \ = \ \ \int_p e^{i p (n + \frac{a}{2})} B^a_\mu(p) \\
 & & \hspace{-0.7cm} \Gamma_{\lambda} \ \ = \ \ \sin (a k_\lambda) \\
 & & \hspace{-0.7cm} c_\mu \ \ = \ \ \cos\left(\frac{a (p+p')_\mu}{2}\right) \quad , \quad s_\mu \ \ = \ \ \sin\left(\frac{a (p+p')_\mu}{2}\right) \\
 & & \hspace{-0.7cm} M_\mu \ \ = \ \ \cos\left(\frac{k_\mu}{2}\right) \quad , \quad N_\mu \ \ = \ \ \sin\left(\frac{k_\mu}{2}\right) \\
 & & \hspace{-0.7cm} W \ \ = \ \ 2 \sum_\lambda \sin^2\left(\frac{k_\lambda}{2}\right) \\
 & & \hspace{-0.7cm} E^2 \ \ = \ \ \sum_{i=1}^3 N^2_i + \frac{a^2 \lambda^2}{4} \quad , \quad E' \ \ = \ \ 2 \textrm{argsh}(E) .
\end{eqnarray}

In Fourier space the action at $\mathcal{O}(g^2_0)$ is given by
\begin{eqnarray}
\nonumber & & \hspace{-0.7cm} S^\textrm{HQET} \ \ = \ \  \int_p \frac{1}{a} Q^\dagger(p) (1 - e^{-i p_4 a}) Q(p) + \delta m Q^\dagger(p) Q(p) \\
\nonumber & & \hspace{0.675cm} + i g_0 \int_p \int_{p'} \int_q \delta(q+p'-p) Q^\dagger(p) B^a_4(q) T^a Q(p') e^{-i (p_4 + p'_4) \frac{a}{2}} \\
\label{actionfourier} & & \hspace{0.675cm} + \frac{a g^2_0}{2!} \int_p \int_{p'} \int_q \int_r \delta(q+r+p'-p) Q^\dagger(p) B^a_4(q) B^b_4(r) T^a T^b Q(p') e^{-i (p_4+p'_4) \frac{a}{2}} .
\end{eqnarray}
The block gauge fields $B^a_\mu$ can be expressed in terms of the usual gauge fields:
\begin{eqnarray}
B_\mu \ \ = \ \ \sum_{n=1}^{\infty} B^{(n)}_\mu ,
\end{eqnarray}
where $B^{(n)}_\mu$ contains $n$ factors of $A$. At next to leading order, it was shown that we only need $B^{(1)}_\mu$ \cite{Lee:2002fj}:
\begin{eqnarray}
 & & \hspace{-0.7cm} B^{(1)}_\mu(k) \ \ = \ \ \sum_\nu h_{\mu \nu}(k) A_\nu(k) \\
 & & \hspace{-0.7cm} h_{\mu\nu}(k) \ \ = \ \ \delta_{\mu \nu} D_\mu(k) + (1 - \delta_{\mu \nu}) G_{\mu \nu}(k) \\
 & & \hspace{-0.7cm} D_\mu(k) \ \ = \ \ 1 - d_1 \sum_{\rho \neq \mu} N^2_\rho +d_2 \sum_{\rho < \sigma, \rho,\sigma \neq \mu} N^2_\rho N^2_\sigma - d_3 N^2_\rho N^2_\sigma N^2_\tau \\
 & & \hspace{-0.7cm} G_{\mu \nu}(k) \ \ = \ \ N_\mu N_\nu \bigg(d_1 - d_2 \frac{N^2_\rho + N^2_\sigma}{2} + d_3 \frac{N^2_\rho N^2_\sigma}{3}\bigg) \\
 & & \hspace{-0.7cm} d_1 \ \ = \ \ \frac{2}{3} \alpha_1 \Big(1 + \alpha_2 (1 + \alpha_3)\Big) \quad , \quad d_2 \ \ = \ \ \frac{4}{3} \alpha_1 \alpha_2 (1 + 2\alpha_3) \quad , \quad d_3 \ \ = \ \ 8 \alpha_1 \alpha_2 \alpha_3 .
\end{eqnarray}

The Feynman rules are the following:

\vspace{0.3cm}
\begin{tabular}{|c|c|}
\hline
 & \vspace{-0.4cm} \\
heavy quark propagator & $a (1-e^{-i p_4 a} + \epsilon)^{-1}$ \\
 & \vspace{-0.4cm} \\
vertex $V^a_{\mu,h h g}(p,p')$ & $-i g_0 T^a \delta_{\mu 4} \sum_\rho h_{\mu \rho} e^{-i (p_4 + p'_4) \frac{a}{2}}$ \\
 & \vspace{-0.4cm} \\
vertex $V^{a b}_{\mu \nu,h h g g}(p,p')$ & $-\frac{1}{2} a g^2_0 \delta_{\mu 4} \delta_{\nu 4} \sum_{\rho,\sigma} h_{\mu \rho} h_{\nu \sigma} \{T^a , T^b\} e^{-i (p_4 + p'_4) \frac{a}{2}}$ \\
 & \vspace{-0.4cm} \\
gluon propagator in the Feynman gauge & $a^2 (C^{-1}_0 \delta_{\mu \nu} \delta^{a b} (2W + a^2 \lambda^2)^{-1} + \Delta_{\mu \nu})$\vspace{-0.4cm} \\
 & \\
\hline
\end{tabular}
\vspace{0.3cm}

Note that $p'$ and $p$ are the in-going and the out-going fermion momenta,
respectively. We also introduce an infrared regulator $\lambda$ for the gluon
propagator. We symmetrize the vertex $V^{a b}_{\mu \nu,h h g g}$ by introducing
the anti-commutator of the $SU(3)$ generators normalized by a factor $1/2$. The
gluon propagator and the vertices are defined with the $A$ field. At one-loop
the infrared regulator to the gluon propagator that we have chosen is
legitimate, because no three-gluon vertex is involved.




\end{document}